\documentstyle[epsfig]{aipproc}
  
\newcommand{\Dal}{{\,\lower 1.0pt\vbox{\hrule \hbox{\vrule height 0.3 cm 
\hskip 0.28 cm \vrule height 0.3 cm}\hrule}\,}}

\begin{document}
\title{One particular approach to the non-equilibrium quantum dynamics}

\author{Eduardo S. Tututi$^{\dagger}$ and Petr Jizba$^*$ }
\address{$^{\dagger}$ECFM, UMSNH, Apdo. Postal 2-71,  58041, 
                 Morelia Mich., M\'exico\thanks{Partially supported by CONACYT 
under grant I-29950 E\,\, 
                 and CIC-UMSNH.}\\
$^*$DAMTP,  University of Cambridge, Silver Street, Cambridge, 
              CB3 9EW, UK\thanks{Fitzwilliam college}} 

\maketitle

\begin{abstract}
We present a particular approach to the non-equilibrium dynamics of quantum
field theory. This approach is based on the Jaynes-Gibbs principle of the 
maximal entropy and its implementation, throughout the initial-value data, into
the dynamical equations for  Green's functions. We use the $\phi^4$ theory in 
the large $N$ limit to show how our method works by calculating the pressure 
for a system which is invariant under both  spatial and  temporal translations.
\end{abstract}


\section*{Introduction: Jaynes-Gibbs principle}

The objective of this talk is to present a novel approach to a non-equilibrium 
dynamics of quantum fields  \cite{jizba-tututi}. This approach is based on the 
Jaynes-Gibbs maximum entropy principle \cite{jaynes}, which, in contrast to 
other approaches in use 
\cite{calzetta-hu,kadanoff-baym,chinesse,eboli-jackiw-pi}, 
constructs a density matrix $\rho$ directly from the  experimental/theoretical
initial-time data (e.g. pressure, density of energy, magnetization, ionization 
rate, etc.). We illustrate our method on the $\phi^4$ theory with the $O(N)$ 
internal symmetry in the large $N$ limit, provided that  the non-equilibrium 
medium in question is translationally invariant.  

To start, we consider the following definition of expectation value of some 
dynamical operator $A$: $\langle A \rangle = {\rm Tr} (\rho A)$, where the
trace is taken with respect to an orthonormal basis of {\em physical} states 
describing the ensemble at some initial time $t_{i}$. Let us consider the 
information (or Shannon) entropy $S[\rho]=-{\rm Tr}(\rho\ln\rho)$ 
\cite{jaynes}. According to the Jaynes-Gibbs principle, we have to maximize 
$S[\rho]$ subject to constrains imposed by  the expectation value  of certain 
experimental/theoretical observables:
$
\langle P_i [\Phi,\partial\Phi] (x_1,\ldots)\rangle = g_i (x_1,\ldots),\quad 
i=1,\ldots n
\, ,
$
where  the operators $P_i [\Phi,\partial\Phi]$, in contrast to thermal 
equilibrium, need not to be constants of the motion; space-time dependences are
allowed.  The maximum of the entropy determines the density matrix with the 
least informative content. 
\begin{eqnarray}
\rho = {1\over {\cal Z}(\lambda_i)} \exp\left(-\sum_{i=1}^n \int 
\prod_j d^4 x_j 
\lambda_i (x_1,\ldots) P_i [\Phi,\partial\Phi]\right)
\, ,
\label{jaynes-gibbs1}
\end{eqnarray}
where $\lambda_i$ are the Lagrange multipliers to be determined. The 
`partition function' ${\cal Z}$ is 
${\cal Z}(\lambda_i) =
{\rm Tr}\left\{\exp\left(-\sum_{i=1}^n \int \prod_j d^4 x_j 
\lambda_i (x_1,\ldots) P_i [\Phi,\partial\Phi]\right)\right\}.$
In the previous equations the time integration is not present at all
(i.e. $g_i(\ldots)$ are  especified at the initial time $t_i$). In case when 
the constraint functions $g_{i}(\ldots)$ are known over some  gathering 
interval 
$(-\tau,t_i)$  the correspondent integration $\int_{-\tau}^{t_{i}} dt$ should 
be present in $\rho$. The Lagrange multipliers $\lambda_i$ might be eliminated 
if one solves $n$ simultaneous equations: 
$g_i=-\delta \ln {\cal Z}/\delta\lambda_i$. 
The solution can be formally written as 
$\lambda_i = \delta S_G[g_1,\ldots , g_n]\mid_{max}/\delta g_i$.

\section*{Off-equilibrium dynamical equations}

In this section we briefly introduce the off-equilibrium dynamical 
(or Dyson-Schwinger) equations. For simplicity we illustrate this on a single 
scalar field $\Phi$ coupled to an external source $J$ described by the  action 
$S'[\Phi]= S[\Phi]+\int J\Phi$.   
Associated with this action we have the functional equation of motion 
\cite{jizba-tututi,chinesse}:
\begin{equation}
{1\over Z[J]}{\delta S\over\delta\Phi }\left[\Phi_{\alpha} =-i 
{\delta\over\delta J} \right]
Z[J]=-J_{\alpha}	
\, ,
\label{off-equilibrium2}
\end{equation}
with $Z[J]={\rm Tr}\{\rho T_C \exp (i\int_C d^4 x J(x)\Phi (x)\}$ being the 
generating functional  of Green's functions. Here $C$ is the Keldysh-Schwinger 
contour  which runs along the real axis  from  $t_i$ to $t_f$ ($t_f > t_i$, 
$t_f$ is arbitrary) and then back to $t_i$.
In (\ref{off-equilibrium2}) we have associated  with the  upper branch of $C$ 
the index $``+"$ and with the lower one the index $``-"$ (in the text we shall 
denote the indices $+/-$ by Greek letters $\alpha, \beta$).

Let us define the classical field $\phi_{\alpha}$ as the expectation value of 
the field operator in the presence of $J$: i.e. 
$\phi_{\alpha}=\langle\Phi_{\alpha}\rangle$. 
Defining the generating functional of the connected Green's functions as 
$Z[J]=\exp (iW[J])$, the two-point Green's function is 
$G_{\alpha\beta}(x,y)=-{\delta^2 W\over \delta J_{\alpha}(x)\delta 
J_{\beta}(y)}=-i\langle T_C\Phi (x)\Phi (y)\rangle+ i\langle\Phi (x)
\rangle\langle \Phi (y)\rangle$. 
Eq.(\ref{off-equilibrium2}) is the first one  of an  infinite hierarchy of 
equations for Green functions. Further equations can be obtained from 
(\ref{off-equilibrium2}) by taking successive variations with respect to $J$. 
True dynamical equations are then obtained if one substitutes the physical 
condition $J=0$ into equations obtained.

To reflect the effects of the density matrix in the Dyson-Schwinger equations 
it is necessary to construct the corresponding boundary conditions.\footnote{
Let us remind that at equilibrium the corresponding boundary conditions are 
the Kubo-Martin-Schwinger (KMS) conditions.} Using the cyclic property of the 
trace  together with the Baker-Campbell-Hausdorff relation: 
$e^A Be^{-A}=\sum_{n=0}^{\infty} {1\over n !}C_n$,
(where $C_0=B$ and $C_n=[A,C_{n-1}]$), and setting $A=\ln (\rho)$ and $B=\Phi 
(x_1)$ with $x_{10}=t_i$ we obtain the generalized KMS conditions:\\
$\langle\Phi (x_1)\cdots\Phi (x_n)\rangle = \langle\Phi (x_2)\cdots\Phi (x_n)
\Phi (x_1) \rangle + \sum_{k=1}^{\infty}{1\over k !}\langle\Phi (x_2)\cdots\Phi
(x_n) C_k (x_1) \rangle\, .$ So namely for the two-point Green function we have
$ G_{+-}(x,y) = G_{-+}(x,y) + \sum_{k=1}^{\infty}{1\over k !}{\rm Tr}\{\rho 
\Phi (x) C_k (x)\}\, .$ As an example of the latter relation we can choose the 
particular situation when $\rho = \exp (-\beta H)/ {\cal{Z}}$, in which case 
we get the well known KMS condition: 
$G_{+-}({\bf x};t, {\bf y};0) = G_{-+}({\bf x};t-i\beta, {\bf y};0)$.

\section*{Example: out-of-equilibrium pressure}

In order to apply our previous results let us consider the $\phi^4$ theory
with the $O(N)$ internal symmetry in the large $N$ limit (also the Hartree-Fock
approximation). It is well known that, in this limit only two-point Green's 
functions are relevant \cite{jizba-tututi,eboli-jackiw-pi,amelinocamelia-pi}.  
The Dyson-Schwinger equations  for $G_{\alpha \beta}$ are automatically 
truncated and reduce to the Kadanoff-Baym equations 
\cite{kadanoff-baym}:
$\left( \Dal + m^2_0+{i\lambda_0\over 2}G_{\alpha\alpha} (x,x) \right)
G_{\alpha\beta} (x,y) = - \delta (x-y)(\sigma_3)_{\alpha\beta}\, ,$
where $\sigma_3$ is the Pauli matrix; $\lambda_0$ and  $m_0$
are, respectively, the bare coupling and the  bare mass of the theory.
If the system is translationally invariant 
the Fourier transform solves the  Kadanoff-Baym equations and the 
correspondingfundamental solution reads:
$ G_{\alpha\beta} (k)  = { (\sigma_3)_{\alpha\beta}\over  k^2+{\cal M}^2 
+ i\epsilon (\sigma_3)_{\alpha\beta} } -2\pi  i \delta ( k^2+{\cal M}^2 )
f_{\alpha\beta} (k) \, , $
where the (finite)  ${\cal{M}}$ is ${\cal{M}}^{2} = m_{0}^{2} + i
\frac{\lambda_{0}}{2}G_{++}(0)$. Function $f_{\alpha\beta}(k)$ must be 
determined through the generalized KMS conditions. 
Let us now choose the constraint to be used. Keeping in mind that we are 
interested in a system which is invariant under both spatial and temporal 
translations, we choose the constraint $g({\bf k}) = \langle \tilde{\cal H} 
({\bf k})\rangle$, where $\tilde{\cal H} = \omega_k a^{\dagger}({\bf k})
a({\bf k})$, with $\omega_k=\sqrt{{\bf k}^2+{\cal M}^2}$ (notice that in the 
large $N$ limit the Hamiltonian is always quadratic in the fields). The 
corresponding density matrix then reads
\begin{eqnarray}
\rho = {1\over {\cal Z}(\beta)} \exp\left(-\int {d^3{\bf k}\over (2\pi)^3 
2\omega_k} \beta ({\bf k}) \tilde{\cal H} ({\bf k})\right)
\, ,
\label{off-equilibrium6}
\end{eqnarray}
with ${\beta ({\bf k})\over (2\pi)^3 2\omega_k}$ being the Lagrange multiplier 
to be determined. According to the maximum entropy principle we find that
$\beta({\bf k})$ fulfils equation
\begin{eqnarray}
g({\bf k})= {V\over (2\pi)^3}{\omega_k\over e^{ \beta ({\bf k}) \omega_k} -1}
\, ,
\label{off-equilibrium7}
\end{eqnarray}
where $V$ denotes the volume of the system. Eq.(\ref{off-equilibrium7}) can be 
interpreted as  the density of energy per mode. Similarly as in the case of 
thermal equilibrium, $\beta ({\bf k})$ could be interpreted as ``temperature" 
with the proviso that different modes have now different ``temperatures".

\begin{figure}[h]
\epsfxsize=6.5cm 
\centerline{\epsffile{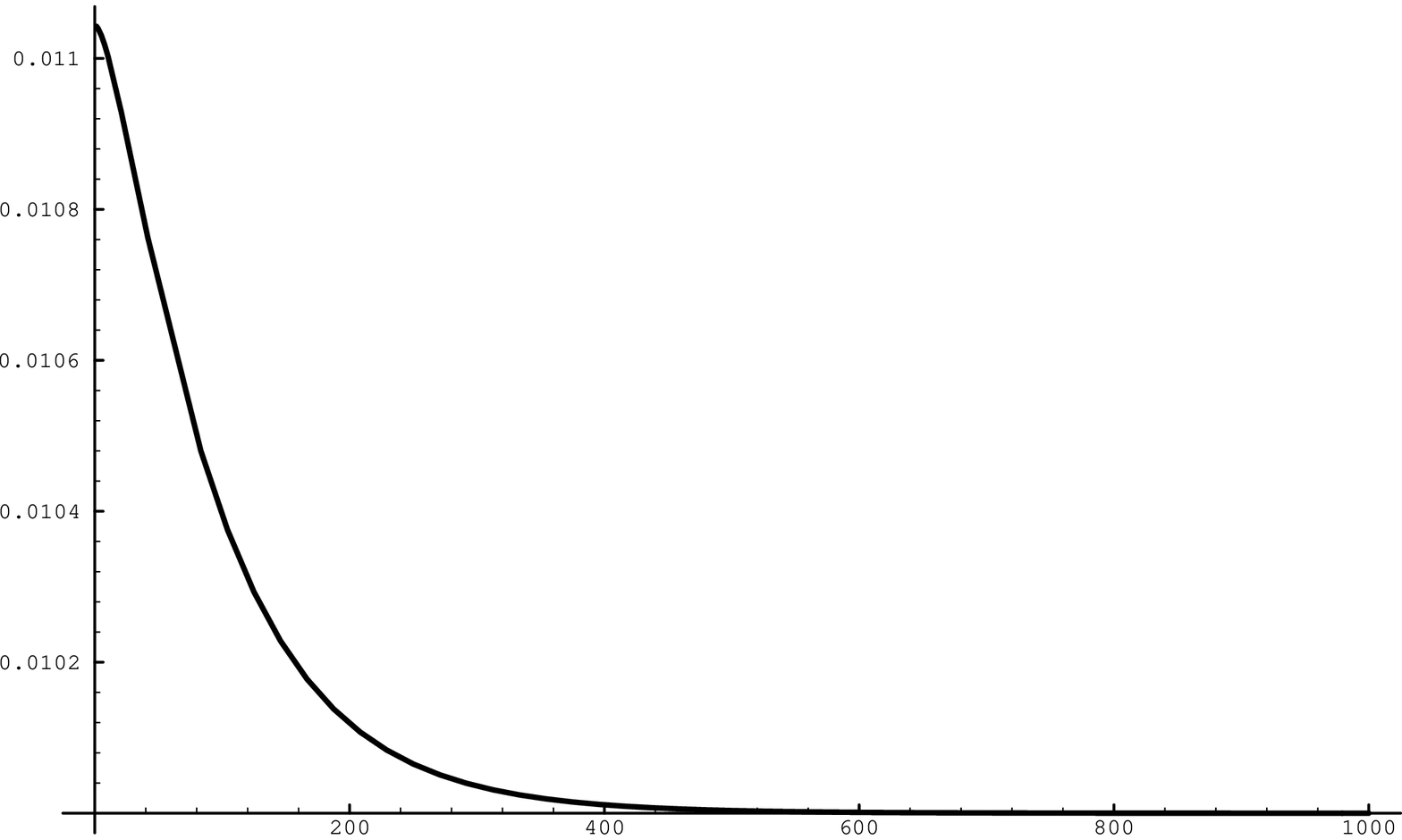}\vspace{1mm}\epsfxsize=6.5cm 
\epsffile{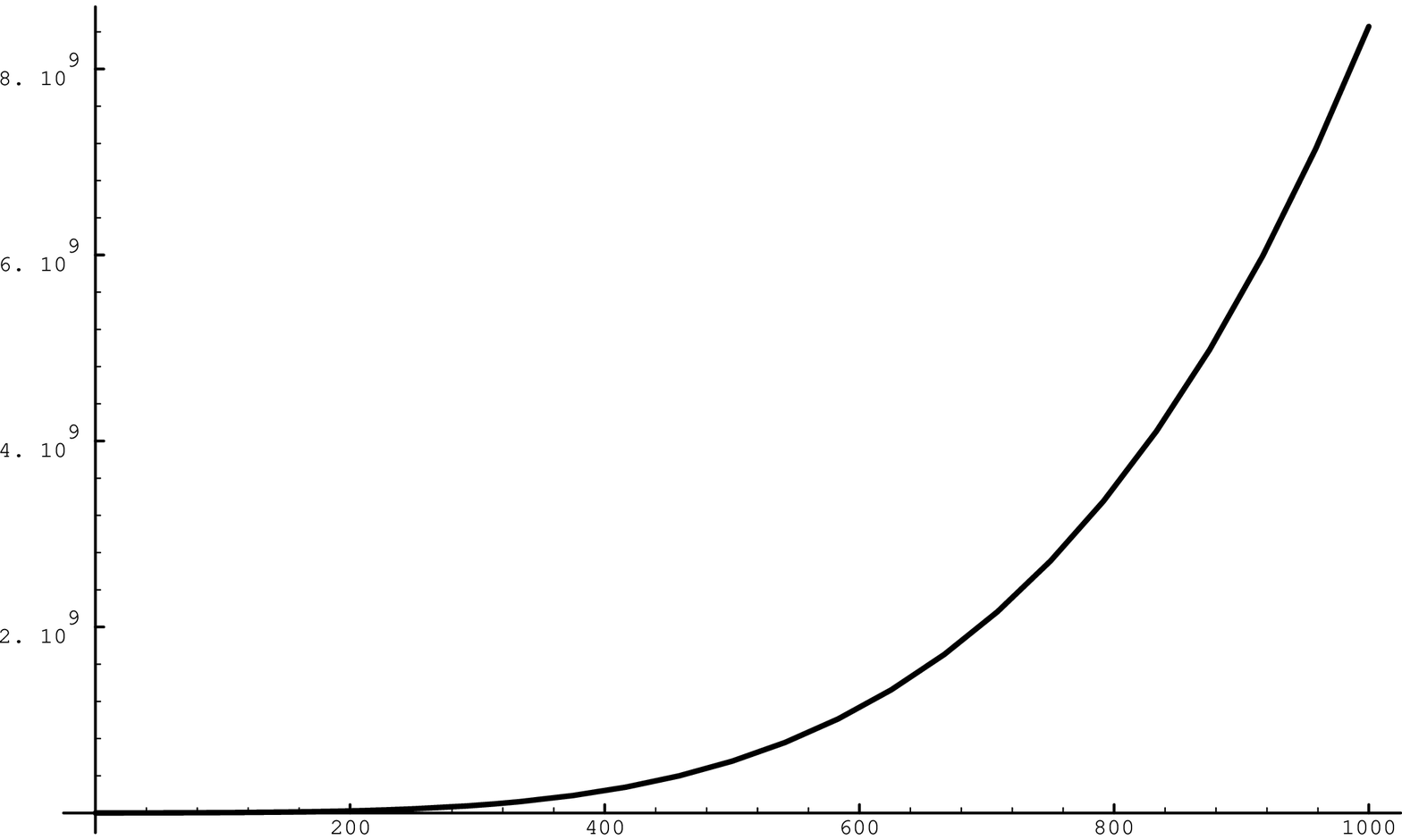}} 
\vspace{-10mm} 
\caption{\em  In $(a)$ we plot Eq.(\ref{off-equilibrium7a}) at $\sigma=100 
Mev$, while in $(b)$ we plot the  difference of   equilibrium    and
non-equilibrium pressures for $m_{r}=100\mbox{MeV}$.} 
\label{fig}
\setlength{\unitlength}{1mm} 
\begin{picture}(10,7)
\put(125,25){\tiny $T, \sigma $  [GeV]}
\put(90,63){\tiny  $P(T)-P(\sigma)[(\mbox{GeV})^{4}]$} 
\put(60,25){\tiny ${\cal M}$ [Mev]}
\put(25,63){\tiny $\langle\beta\rangle$ [Mev]}
\put(105,20){\tiny $(b)$}
\put(35,20){\tiny $(a)$}
\end{picture}
\end{figure}

The generalised KMS conditions in this case are $G_{+-}(k) = 
e^{-\beta({\bf k})k_{0}}G_{-+}(k)$,  and so the corresponding $f_{++}$ reads: 
$f_{++} = [\exp (\beta ({\bf k}) \omega_k) -1]^{-1}$.  Let us now consider a 
particular system in which $g({\bf k})= {V\over (2\pi)^3}\exp 
(\omega_k/\sigma)$. In this case $\sigma$ is the physical parameter which, as 
we shall see below, can be interpreted as a ``temperature'' parameter.  This 
particular choice corresponds to a system where the lowest frequency modes 
depart  from equilibrium, while the high energy ones obey the Bose-Einstein 
distribution (typical situation  in many non-equilibrium media, e.g. plasma 
heated up by ultrasound waves, hot fusion or ionosphere ionised by sun). In 
terms of the  parameter  $\sigma$ the Lagrange multiplier may be written as  
$\beta ({\bf k})= {1\over\sigma}+{1\over\omega_k}\sum_{n=1}^{\infty}
{(-1)^{n+1}\over n}\exp ( -n\omega_k/\sigma)$. Notice that when $\omega_k
\gg\sigma$, $\beta\sim\sigma^{-1}$, and we may see that $f_{++}$ approaches 
to the  Bose-Einstein distribution with temperature $\sigma$. However, when 
$\omega_k\sim\sigma$ the latter interpretation fails.
Instead of the parameter $\sigma$, it may be useful to work with the 
expectation value of $\beta ({\bf k})$:
\begin{equation}
\langle\beta\rangle   =  {\int    d^3{\bf    k}\,   \beta({\bf  k})e^{
-{\omega_{k}/\sigma}}\over     \int d^3{\bf k} \,e^{-{\omega_k/\sigma}}}
=  {1\over    \sigma}    +  {    {\sum}_{n=1}^{\infty}
{(-1)^{n+1}\over    n(n+1)}K_1({(n+1){\cal{M}}/\sigma})\over {\cal{M}}
K_2({{\cal{M}}/\sigma})} \, , 
\label{off-equilibrium7a} 
\end{equation}
where $K_n$ is the Bessel function of imaginary argument of order $n$. An 
interesting feature of Eq.(\ref{off-equilibrium7a}) is that it is actually 
insensitive to the value of ${\cal M}$ which is important if one wants to use 
$1/\langle \beta \rangle$ as a ``temperature''. The actual behaviour of 
$\langle \beta \rangle$ is depicted in Fig.\ref{fig}
\vspace{-25mm} 

Let us  now consider the  renormalized expression for the  expectation value 
of the energy momentum tensor \cite{jizba-tututi}:
\begin{displaymath}
\langle\theta_{\mu\nu}\rangle_{\rm ren} =N \int {d^dk\over 
(2\pi)^d}k_{\mu}k_{\nu} [G_{++}(k) -G(k)] -i{N g_{\mu\nu}\delta m^2\over 4} 
\int {d^dk\over (2\pi)^d} [G_{++}(k) +G(k)]
\, ,
\end{displaymath}
with $G$ being the usual ($T=0$) causal Green function and $\delta m^2 = 
{\cal M}^2 -m^2_{r}$ with $m_{r}$ being the ($T=0$) renormalized mass. The 
pressure per particle, in the high ``temperature" expansion (i.e. for large 
$\sigma$ or small $\langle \beta \rangle$) for the system  described by the 
density matrix (\ref{off-equilibrium6}) may be worked out either in terms 
of $\sigma$, using the Mellin transform technique 
\cite{jizba-tututi,landsman-weert}:
\begin{displaymath}
P(\sigma)=-{1\over 3N}\langle\theta ^i_{\;i}\rangle_{\rm ren}= 
\frac{\sigma^{4}}{\pi^{2}} -
\frac{\sigma^{2}\, {\cal{M}}^{2}}{2\pi^{2}}+ \frac{\lambda_{r}}{8}\left(
\frac{\sigma^{2}{\mathcal{M}}^{2}}{64  \pi^{4}}  - \frac{\sigma^{4} 3}{4
\pi^{4}} \right)  + {\mathcal{O}}\left({\mbox{ln}}({\mathcal{M}}/\sigma);  
\lambda_{r}^{2}\right)\, ,
\end{displaymath}
or in terms of $1/\langle \beta \rangle$ using the Pad{\'e} 
approximation\cite{jizba-tututi}:
\begin{eqnarray*} 
P(\langle \beta  \rangle )  &=& 0.0681122\,{\langle \beta  \rangle^{-4}}
-0.0415368\, {\langle \beta \rangle^{-2}}{\cal{M}}^{2}\ + \lambda_{r}\,
\left( -0.000647\,{\langle  \beta   \rangle^{-4}} \right.\\ \nonumber  &+& 
\left. 0.0000164\,{\langle \beta \rangle^{-2}}\,{\mathcal{M}}^{2}\right) 
+  {\mathcal{O}}\left({\mathcal{M}}^{2} {\mbox{ln}}({\mathcal{M}}\langle 
\beta\rangle); \lambda_{r}^{2} \right).
\end{eqnarray*}
It is interesting to compare the previous two results with the high-temperature
expansion of the same system in thermal equilibrium  \cite{amelinocamelia-pi}:
\begin{displaymath}
P(T) = \frac{T^{4}\;  \pi^{2}}{90} - \frac{T^{2}\;{\cal M}^2}{24}  +
\frac{T\;{\cal M}^{3}}{12  \pi}+ \frac{\lambda_{r}}{8}\left(\frac{T^{4}}{144}-
\frac{T^{3}\;{\cal M}}{24   \pi}   +       \frac{T^{2}\;{\cal M}^2}{16
\pi^{2}}\right)
+  {\cal{O}}\left( \mbox{ln}\left(  \frac{{\cal M}}{T4\pi
}\right)\right).
\end{displaymath}
Particularly, the leading ``temperature'' coefficients in the first two 
expansions approximate to a very good accuracy the usual Stefan-Boltzmann 
constant for scalar theory.  The latter vindicates the interpretation of 
$\sigma$ and $1/\langle \beta \rangle$ as  temperatures for high energy modes. 
The behaviour of both $P(T)$ and $P(\sigma)$ are shown in Fig.\ref{fig}.

\section*{summary and outlook}

One  of the main advantages of the Jaynes-Gibbs construction is that one starts
with constraints imposed by experiment/theory. The constraints directly 
determine the density matrix with the least informative content (the least 
prejudiced density matrix which is compatible with all information one has 
about the system) and consequently the generalized KMS conditions for the  
Dyson-Schwinger equations. We applied our method on a toy model system 
($O(N)\; \lambda\phi^{4}$ theory), in the translationally invariant medium.
The method presented, however, has a natural potential to be extensible to more
general systems. Particularly to media where the translational invariance is 
lost. As an example we can mention systems which are in local thermal 
equilibrium. For such systems it is well known\cite{jaynes,landsman-weert} 
that equilibrium $\beta$ must be replaced by $\beta ({\bf x})$ 
(i.e. temperature which slowly varies with position). Obviously one may receive
this result from the outlined Jaynes--Gibbs principle almost for free. Work 
on more complex systems is now in progress.

\end{document}